\documentclass{scrartcl}

\usepackage[affil-it]{authblk}

\usepackage{hyperref}

\usepackage{amsfonts}
\usepackage{amssymb,amsmath}
\usepackage{amsthm}

\newtheorem{theorem}{Theorem}
         
\newtheorem{lemma}{Lemma}

\usepackage{cite}

\newcommand{\reals}{\ensuremath{\mathbb{R}}}
\newcommand{\complexes}{\ensuremath{\mathbb{C}}}

\DeclareMathOperator{\E}{\mathbb{E}}
\DeclareMathOperator{\var}{Var}

\begin{document}


\title{Note on Closed-Form Expressions for Irreducible Loop Integrals
}

\author{Richard Chapling\thanks{Electronic address: 
\url{rc476@cam.ac.uk}}}

\affil{Department of Applied Mathematics and Theoretical Physics, \\ University of Cambridge, Cambridge, England}

\maketitle


\begin{abstract}
	We provide a new analysis of the irreducible loop integrals first considered by Wu  \cite{ISI:000187232400003}. Using convergence ideas from probability, we produce conditions on the regulator masses so that the integrals have well-defined limits in the limit required by the regularisation technique; we then derive closed-form expressions for the regularised forms of these integrals in terms of incomplete Gamma-functions.
\end{abstract}

\section{Introduction}

In a recent paper \cite{Chapling:2016LORE} we proved three conjectures of Wu \cite{ISI:000187232400003} required in the method of Loop Regularisation. In this note we give a direct analysis of the irreducible loop integrals (ILIs) involved in the formulation of the theory; this enables us to produce closed-form expressions for the limits required in loop regularisation.

The second section contains an outline of the theory of Loop Regularisation, focussing on the introduction and properties of the ILIs. In the third, we summarise the original method of evaluating the integrals, indicating the source of the conjectures proven in our previous paper \cite{Chapling:2016LORE}, but also deriving the form of the integral that we consider in the final section, namely
\begin{equation}
	\lim_{\substack{ m \to \infty \\ M_{R}^{2} \to \infty }} \int_{0}^{\infty} \tau^{\alpha-1} e^{-\tau \mu^{2}} \left( 1-e^{-\tau M_{R}^{2}} \right)^{m} d\tau.
\end{equation}
Wu discusses this as an outcome of the \emph{proper-time formalism} in \cite{ISI:000224011700003}. We first consider some properties of the bracketed factor, which then allow us to use ideas from probability theory to understand its behaviour in the large-\(m\) limit: armed with this, we may deduce natural conditions on the way we can take \(M_{R} \to \infty\) while ensuring the finiteness of this integral.

Calculations shall be carried out in \(d\) dimensions; it will become apparent that with judicious choice of the quantities involved, the actual number of dimensions is not an essential feature of the theory as we investigate it here, although of course it is significant in applications. The metric \(g_{\mu\nu}\) is Minkowski with signature \( (+,-,-,-) \). Our notation is consistent with \cite{Chapling:2016LORE}, but not necessarily with \cite{ISI:000187232400003}.

\section{Loop Regularisation}

We outline briefly the details of a specific regularisation scheme known as Loop Regularisation, which was first discussed by Wu in a paper of 2003, \cite{ISI:000187232400003}; we follow loosely parts of that paper in our presentation in this section.

Initially, the second section of Wu's paper outlines a new set of conditions for a regularisation scheme to preserve gauge invariance. We consider the following set of integrals, known as \emph{irreducible loop integrals} or ILIs:
\begin{align*}
	I_{-2\alpha} &= \frac{1}{(2\pi)^{d}}\int_{\reals^{d}} \frac{dp}{(p^{2}-\Delta)^{d/2+\alpha}} \\
	I_{-2\alpha \, \mu\nu} &= \frac{1}{(2\pi)^{d}}\int_{\reals^{d}} \frac{p_{\mu}p_{\nu}}{(p^{2}-\Delta)^{d/2+1+\alpha}} \, dp \\
	I_{-2\alpha \, \mu\nu\rho\sigma} &= \frac{1}{(2\pi)^{d}}\int_{\reals^{d}} \frac{p_{\mu}p_{\nu}p_{\rho}p_{\sigma}}{(p^{2}-\Delta)^{d/2+2+\alpha}} \, dp.
\end{align*}
Wu shows by application of Ward identities and calculation of the Yang--Mills vacuum diagrams that the regularised versions of these, \( I^{R}_{2} \) \&c., should satisfy the following identities:
\begin{enumerate}
\item
Quadratically divergent:
\begin{align*}
	I^{R}_{2 \, \mu\nu} &= \frac{1}{2} g_{\mu\nu} I^{R}_{2}, \\
	I^{R}_{2 \, \mu\nu\rho\sigma} &= \frac{1}{8} (g_{\mu\nu}g_{\rho\sigma} + g_{\mu\rho}g_{\sigma\nu} + g_{\mu\sigma}g_{\nu\rho}) I^{R}_{2},
\end{align*}

\item
Logarithmically divergent:
\begin{align*}
	I^{R}_{0 \,\mu\nu} &= \frac{1}{4} g_{\mu\nu} I^{R}_{2}, \\
	I^{R}_{0 \, \mu\nu\rho\sigma} &= \frac{1}{24} (g_{\mu\nu}g_{\rho\sigma} + g_{\mu\rho}g_{\sigma\nu} + g_{\mu\sigma}g_{\nu\rho}) I^{R}_{0}
\end{align*}
\end{enumerate}

(We observe later that dimensional regularisation satisfies this, for example.)

The remainder of Wu's paper is dedicated to producing a regularisation scheme (\emph{Loop Regularisation}, or LORE) consistent with these, that operates strictly in four dimensions, and can also maintain non-Abelian gauge invariance.\footnote{Unlike Pauli--Villars regularisation, for example: \cite{Dai:2005ve,ISI:000258778900002,ISI:000293182400002}. For further details of the advantages and consequences of LORE, see Wu's review article \cite{Wu:2014uq}.}

The method is as follows:
\begin{enumerate}
\item
Introduce a set of \(m\) regulating masses \(M_{k}\), and replace the integrated loop momenta \( p^{2} \) by \( [p^{2}]_{k} := p^{2}+M_{k}^{2} \)
\item
Replace the loop integral by a regularised measure,
\begin{equation*}
	\int dp \mapsto \int [dp]_{k} := \lim_{m \to \infty} \lim_{M_{k}^{2} \to \infty} \sum_{k=0}^{m} c^{m}_{k} \int dp,
\end{equation*}
where the \(c_{k}^{m}\) are constants to be determined.
\item
Calculate the integrals with a finite number of finite regulating masses
\item
Take the number of regulating masses, and their magnitudes, to \( \infty \), in a way specified below.
\end{enumerate}

One recovers the original integral from taking \( M_{0}=0 \), \( c^{m}_{0}=1 \), and the rest of the \(M_{l}\) to \( \infty \).  If the original integrals are IR divergent, one may prefer to take \( M_{0} = \mu_{s} \), some small cutoff, to avoid this. However, the key idea of this regularisation is that we can get rid of the slowly-decaying terms at \( p^{2} \to \infty \): notice that for large \(p^{2}\),
\begin{align*}
	\sum_{k=0}^{m} c^{m}_{k} \frac{1}{(p^{2}+M_{k}^{2})^{\beta}} &= \sum_{k=0}^{m} c^{m}_{k} \sum_{n=0}^{\infty} \frac{1}{(p^{2})^{\beta}} \binom{\beta}{n} \left( \frac{M_{k}^{2}}{p^{2}} \right)^{n} \\
	&= \sum_{n=0}^{\infty} \binom{\beta}{n} \frac{1}{(p^{2})^{\beta+n}}  \sum_{k=0}^{m} c^{m}_{k} ( M_{k}^{2})^{n}.
\end{align*}
Therefore if we take \(m\) sufficiently large, we are able to eliminate the ultraviolet divergence of an integrand by removing as many of the low-\(n\) powers of \(1/p^{2}\) as necessary, by choosing \(M_{k}\) and \(c_{k}^{m}\) to satisfy the equations\footnote{In fact, this only needs to be satisfied in the limit as the \(M_{k}\) become large, but we can demand it identically to save ourselves from having to change the order of limits more often than necessary.}
\begin{equation*}
	\sum_{k=0}^{m} c_{k}^{m} (M_{k}^{2})^{n} = 0, \qquad c^{m}_{0}=1.
\end{equation*}
Probably the simplest choice for \( M_{k}^{2} \) is to allow it to depend linearly on \(k\), so that \( M_{k}^{2} = \mu_{s}^{2}+k M_{R}^{2} \), where \(M_{R}\) is an overall mass that we can take to \( \infty\) in some manner later. Indeed, it will be necessary to take both \(m \to \infty\) and \( M_{R} \to \infty \) to eliminate any explicit dependence on the specific nature of the regularisation. With this choice, the above constraints become
\begin{equation}
\label{eq:ckmk^nrelation}
	\sum_{k=0}^{m} c_{k}^{m} k^{n} = 0, \qquad c^{m}_{0}=1
\end{equation}
by expanding out, and we recognise that a solution is provided by \( c_{k}^{m} = (-1)^{k} \binom{m}{k} \) (indeed, it is the unique solution).

The crux of the matter now becomes to evaluate the ILIs and demonstrate that the consistency conditions are satisfied; the evaluation also gives useful expressions for calculating amplitudes, of course.

\section{Evaluation of the Integrals Using Sums}

\subsection{Evaluation of the scalar integrals}
\label{sec:scalarinteval}

Scalar integrals are in general simpler to evaluate, so we we give some details of the calculation. The tensor ones are done by using tensor relationships, which may be applied to convergent integrals, as is shown in the next section.

The main calculation proceeds as follows: after Wick rotating and using Schwinger's parametrisation, we have
\begin{equation}
	\frac{1}{(p^{2}+M'^{2}_{k})^{d/2+\alpha}} = \frac{1}{\Gamma(d/2+\alpha)}\int_{0}^{\infty} \tau^{d/2+\alpha-1} e^{-\tau(p^{2}+M'^{2}_{k})} d\tau,
\end{equation}
where we write \( M'^{2}_{k} := \Delta+M^{2}_{k} =: \mu^{2}+k M^{2}_{R} \). Interchanging the order of integration and using the Gaussian integral \( \int_{\reals^{d}} e^{-\lambda p^{2}} \, dp = (\pi/\lambda)^{d/2} \), we have
\begin{align*}
	I^{R}_{-2\alpha}(m) &= \frac{i(-1)^{-\alpha-d/2}}{(4\pi)^{d/2}\Gamma(d/2+\alpha)}\int_{0}^{\infty} \tau^{\alpha-1} \left( \sum_{k=0}^{m} (-1)^{k} \binom{m}{k} e^{-\tau M'^{2}_{k}} \right) d\tau \\
	&= \frac{i(-1)^{-\alpha-d/2}}{(4\pi)^{d/2}\Gamma(d/2+\alpha)}\int_{0}^{\infty} \tau^{\alpha-1} e^{-\tau \mu^{2}} \left( \sum_{k=0}^{m} (-1)^{k} \binom{m}{k} e^{-\tau k M^{2}_{R}} \right) d\tau
\end{align*}
The constraints being satisfied, we can see using series expansions and \eqref{eq:ckmk^nrelation} that the bracket is of the order of \( \tau^{m} \) as \(\tau \downarrow 0\), so we can treat the integral as convergent, even if \( \alpha<0\). Hence it is continuous in \(\alpha\), and we can take limits. Evaluating the integral gives
\begin{equation*}
	I^{R}_{-2\alpha}(m) = \frac{i(-1)^{-\alpha-d/2}}{(4\pi)^{d/2}}\frac{ \Gamma(\alpha)}{\Gamma(d/2+\alpha)} \sum_{k=0}^{m} (-1)^{k} \binom{m}{k} (M'^{2}_{k})^{-\alpha}
\end{equation*}
when \( \alpha>-m \) and is not a negative integer. If \(\alpha\) is a nonpositive integer, we take the limit of \( \Gamma(\alpha) \) times the sum, which gives a derivative with respect to \(\alpha\) due to the pole, but we shall deal with this shortly; the significant point is that this identity, correctly interpreted, holds for \(m\) finite and sufficiently large.

\subsection{Evaluation of the tensor integrals}

When the integral is convergent, we can apply the principle
\begin{equation}
	\int_{\reals^{d}} p_{\mu}p_{\nu} F(p^{2}) \, dp = \frac{1}{d} g_{\mu\nu} \int_{\reals^{d}} p^{2} F(p^{2}) \, dp
\end{equation}
for isotropic integrals to compute the integral directly. Then we apply \( \int_{\reals^{d}} p^{2} e^{-\lambda p^{2}} \, dp = \pi^{d/2}\lambda^{-d/2-1}d/2 \), so
\begin{equation*}
	I^{R}_{-2\alpha \, \mu\nu}(m) = g_{\mu\nu} \frac{i(-1)^{-\alpha-d/2}}{(4\pi)^{d/2}} \frac{\Gamma(\alpha)}{2\Gamma(d/2+\alpha+1)} \sum_{k=0}^{m} (-1)^{k} \binom{m}{k} (M'^{2}_{k})^{-\alpha},
\end{equation*}
with the same provisos regarding \( \alpha>-m \) and the nonpositive integers.

We now see that
\begin{equation}
	I^{R}_{-2\alpha \, \mu\nu}(m) = \frac{1}{d+2\alpha} g_{\mu\nu} I^{R}_{-2\alpha}(m),
\end{equation}
and putting \( d=4 \) and \(\alpha=-1,0\) gives the identities required. The rank-four integrals are similar.

\subsection{Evaluation of the sums and the interesting functional limits}

We now give a brief outline of the large-\(m\) expansions which require the results proven in \cite{Chapling:2016LORE}; these originate from Wu's requirements for the convergence of certain sums obtained in evaluating the limit of the regularised integrals. This is mainly by way of contrast with the method we consider in the next section, where the integral is treated without using series expansions.

We wish to evaluate the function
\begin{equation*}
	\Gamma(\alpha) \sum_{k=0}^{m} (-1)^{k} \binom{m}{k} (M'^{2}_{k})^{-\alpha},
\end{equation*}
to find the explicit values of the regularised integrals. The main question is how to take the limit \( M_{R}^{2} \to \infty \) to obtain finite, sensible quantities; a simple way to consider this is to examine this sum for the quadratically divergent \( I^{R}_{2} \), i.e. with \(\alpha=-1\), when it takes on the form
\begin{align*}
	\sum_{k=0}^{m} & (-1)^{k} \binom{m}{k} (\mu^{2}+k M_{R}^{2}) \log{(\mu^{2}+k M_{R}^{2})} \\
	&= M_{R}^{2} \sum_{k=1}^{m} (-1)^{k} \binom{m}{k} k \log{k} +
	M_{R}^{2} \sum_{k=1}^{m} (-1)^{k} \binom{m}{k} k \log{\left(1+\frac{\mu^{2}}{k M_{R}^{2}}\right)} \\
	&\quad + \mu^{2} \log{\left(\frac{\mu^{2}}{M_{R}^{2}}\right)} + \mu^{2} \sum_{k=1}^{m} (-1)^{k} \binom{m}{k} \left(\log{k} + \log{\left(1+\frac{\mu^{2}}{k M_{R}^{2}}\right)} \right),
\end{align*}
after using \( \sum_{k=0}^{m} (-1)^{k} \binom{m}{k} (a+bk) = 0 \). We shall certainly require the first term to be finite in the limit, i.e. there is a constant \(M_{c}^{2}\) so that
\begin{equation}
	M_{R}^{2} \sum_{k=0}^{m} (-1)^{k} \binom{m}{k} k \log{k} \to M_{c}^{2} \quad \text{as } M_{R}^{2}, m \to \infty.
\end{equation}
This is the source of the first conjecture: on numerical evidence, Wu hypothesises that we should take
\begin{equation*}
	M_{R}^{2} = M_{c}^{2} \log{m};
\end{equation*}
this choice is shown to be correct by our \cite{Chapling:2016LORE}
\begin{theorem}
	Let $m$ be a positive integer. Then
	\begin{align}
	\label{eq:conj1}
		\sum_{k \geqslant 1} (-1)^{k}\binom{m}{k} k \log{k} \sim \frac{1}{\log{m}} \quad \text{as } m \to \infty.
	\end{align}
\end{theorem}

Of course, we then have to take this into account when taking the limit of the other expressions found above. The next term to worry about is also included in the \( \alpha=0 \) sum, which is
\begin{align*}
	\sum_{k=0}^{m} & (-1)^{k} \binom{m}{k} \log{(\mu^{2}+k M_{R}^{2})} \\
	&= \log{\left(\frac{\mu^{2}}{M_{c}^{2}}\right)} + \left( \sum_{k=1}^{m} (-1)^{k} \binom{m}{k} \log{k} -\log{\log{m}} \right) + \sum_{k=1}^{m} (-1)^{k} \binom{m}{k}\log{\left(1+\frac{\mu^{2}}{k M_{R}^{2}}\right)}.
\end{align*}
For consistency, the middle bracket has to converge to a constant; this is resolved by
\begin{theorem}
	Let $m$ be a positive integer. Then
	\begin{align}
	\label{eq:conj2}
		\sum_{k \geqslant 1} (-1)^{k}\binom{m}{k} \log{k} \sim \log{\log{m}} + \gamma \quad \text{as } m \to \infty,
	\end{align}
	where $\gamma$ is the Euler--Mascheroni constant.
\end{theorem}

Lastly, in expanding the logarithm and binomial terms, one has to find limits of sums of the form
\begin{equation*}
	n!\sum_{k=1}^{m} (-1)^{k-1} \binom{m}{k} \frac{1}{(k\log{m})^{n}}
\end{equation*}
as \(m \to \infty\), which is a consequence of
\begin{theorem}
	Let $m,n$ be positive integers. Then
	\begin{align}
	\label{eq:conj3}
		\sum_{k \geqslant 1} (-1)^{k-1}\binom{m}{k} \frac{1}{k^{n}} \sim \frac{\log^{n}{m}}{n!} \quad \text{as } m \to \infty.
	\end{align}
\end{theorem}

It is apparent that this method of treating the integral, while a source of some interesting mathematical results, is a rather dissatisfactory method of producing the evaluation of the limit, and gives inelegant expressions in terms of series, where much care has to be taken with the interchange of the various limiting operations. In the next section, we avoid this by returning to the integral and evaluating the limit without expanding in sums.

\section{A New Approach to Evaluating the ILIs in Terms of Special Functions}

In this section we shall discuss an alternative approach, avoiding introducing any more summations than strictly necessary. We consider the integral
\begin{equation}
	\int_{0}^{\infty} \tau^{\alpha-1} e^{-\tau \mu^{2}} \left( \sum_{k=0}^{m} (-1)^{k} \binom{m}{k} e^{-\tau k M^{2}_{R}} \right) d\tau \\
	= \int_{0}^{\infty} \tau^{\alpha-1} e^{-\tau \mu^{2}} \left( 1-e^{-\tau M_{R}^{2}} \right)^{m} d\tau,
\end{equation}
from the expression for \(I^{R}_{-2\alpha}\) that we found in \S~\ref{sec:scalarinteval}. Changing to the non-dimensional variable \( u=\tau \mu^{2} \), we have to understand the limit of
\begin{equation}
\label{eq:nondimlintegral}
	(\mu^{2})^{-\alpha} \int_{0}^{\infty} u^{\alpha-1} e^{-u} \left( 1-e^{- f(m) u} \right)^{m} du,
\end{equation}
as \(m \to \infty\), where \( f(m):= M_{R}^{2}/\mu^{2} \). Notably, if \( \alpha \leq 0 \), the singularity in \(u^{\alpha-1}\) is cancelled by the zero in the factor \( \left( 1-e^{- f(m) u} \right)^{m} \) providing that \(m\) is large enough; this is corresponds to the regularisation of the ultraviolet divergence that the regularisation provides in the original form of the integral.

Therefore the key lies in analysing the factor \( F_{m}(u) = ( 1-e^{- f(m) u } )^{m} \), and its behaviour as \(m \to \infty\). We first note the following properties:
\begin{enumerate}
\item
\(F_{m}(u)\) is positive on \( (0,\infty) \).
\item
\( F_{m}(u) = O(u^{m}) \) as \( u \downarrow 0 \).
\item
\( F_{m}(u) \to 1 \) as \( u \to \infty \).
\item
\( F_{m}'(u) = m f(m) e^{-f(m)u} (1-e^{-f(m)u})^{m-1} > 0 \) on \( (0,\infty) \).
\end{enumerate}

Therefore \(F_{m}\) does behave as a smooth cutoff at \(0\). Given that \(0\) is precisely the point we wish to avoid, we need to make sure that there is an interval \( (0,\varepsilon) \) in which \( F_{m}(u) \to 0 \) as \( m \to \infty \). Fundamentally this comes down to the position and width of the transition of \( F_{m} \) from \(0\) to \(1\), and we need to choose \(f(m)\) to affect this correctly as \( m \to \infty \). It is not obvious from the above how to determine the correct behaviour.

However, there is another way to approach this problem, from which emerges a clear solution. Namely, it is also possible to consider \(F_{m}\) as the cumulative distribution function of a random variable \(X_{m}\) on \( (0,\infty) \). With this perspective, it is natural to ask about the expectation and variance of \(X_{m}\), which will tell us about the location and shape of \(F_{m}\)'s transition from \(0\) to \(1\). We find that
\begin{align*}
	\E[X_{m}^{n}] &= \int_{0}^{\infty} nu^{n-1} F_{m}(u) \, du \\
	&= \int_{0}^{\infty} nu^{n-1}  (1-e^{-f(m)u})^{m} du \\
	&= \frac{n!}{f(m)^{n}} \sum_{k=1}^{m} (-1)^{k-1} \binom{m}{k} \frac{1}{k^{n}}.
\end{align*}
Hence all moments exist, and by the results in \S~4 of \cite{Chapling:2016LORE}, we can calculate the first two explicitly to obtain
\begin{align*}
	\E[X_{m}] &= \frac{H_{m}^{(1)}}{f(m)}, \\
	\E[X_{m}^{2}] &=  \frac{H_{m}^{(2)}}{f(m)^{2}} + \frac{(H_{m}^{(1)})^{2}}{f(m)^{2}},
\end{align*}
\( H_{m}^{(n)} := \sum_{k=1}^{m} k^{-n} \) being the harmonic number of order \(n\). Thus the variance is 
\begin{equation*}
	\var(X_{m}) = \E[X_{m}^{2}]-(\E[X_{m}])^{2} = \frac{H_{m}^{(2)}}{f(m)^{2}};
\end{equation*}
since \( H_{m}^{(2)} < \pi^{2}/6 \) is bounded and \( f(m) \to \infty \), we conclude that as \(m \to \infty\), the variance tends to zero.

We now apply some results of analysis, phrased in the language of probability: suppose that \begin{equation}
	z = \lim_{m \to \infty} \E[X_{m}] = \lim_{m \to \infty} \frac{H_{m}^{(1)}}{f(m)}
\end{equation}
exists. If we let \(X\) be a random variable that takes on the value \(z\) with probability \(1\), then \( \E[(X_{m}-X)^{2}] \to 0 \), so \(X_{m} \to X\) in mean square, and it follows that \(X_{m} \to X \) in distribution.\footnote{\cite{GrimmettStirzaker:2001pr}, \S~7.2, p.~310, Theorem (4) \emph{et seq}.} This means that, \( F_{m} \to F \), where \( F=\chi_{[z,\infty)} \) is the distribution function of \(X\).

We can now return to the integral we actually wish to evaluate. We have shown that \( F_{m} \to F \) pointwise. We also know that if (and only if) we also have \(z>0\), for sufficiently large \(m\), \(u^{\alpha-1} e^{-u} F_{m}(u)\) is summable and bounded by a constant multiple of \( e^{-u} \). Then we may use the Dominated Convergence Theorem to pass the limit inside the integral sign of the expression in \eqref{eq:nondimlintegral}, obtaining
\begin{equation*}
	\lim_{m \to \infty} \int_{0}^{\infty} u^{\alpha-1} e^{-u} F_{m}(u) du = \int_{z}^{\infty} u^{\alpha-1} e^{-u} \, du = \Gamma(\alpha,z),
\end{equation*}
the standard special function known as the \emph{upper incomplete Gamma-function}.

To summarise, to make this integral well-defined and finite, we require that \(z\) exists and is greater than zero; this occurs, essentially by definition, if \( f(m) \) is asymptotic to \( \log{m} \) as \(m \to \infty \). We can most easily achieve this by putting \( f(m) = z^{-1} \log{m} \). Translating back to physical variables, we see that this has naturally lead to us taking Wu's choice (\cite{ISI:000187232400003}, equation (4.1))
\begin{equation}
	M_{R}^{2} = M_{c}^{2} \log{m}.
\end{equation}
With this, the loop integral becomes
\begin{equation}
	I^{R}_{-2\alpha} = \frac{i(-1)^{-\alpha-d/2}}{(4\pi)^{d/2}\Gamma(d/2+\alpha)} (\mu^{2})^{-\alpha} \Gamma\left(\alpha, \frac{\mu^{2}}{M_{c}^{2}} \right).
\end{equation}

One can check that this agrees with the explicit expressions given by Wu for \(I^{R}_{2}\), \(I^{R}_{0}\) and \(I^{R}_{-2}\) ((3.22--24) in \cite{ISI:000187232400003}). They also clearly exhibit the singularities typical of loop regularisation: the incomplete Gamma-function is given by
\begin{equation}
	\Gamma(\alpha,z) = \int_{z}^{1}  u^{\alpha-1} e^{-u} \, du + \int_{1}^{\infty} u^{\alpha-1} e^{-u} \, du;
\end{equation}
the latter is a constant, and the former can be expanded in what may be called a Frobenius series, with most singular term \( -z^{\alpha}/\alpha \), which is, for example, quadratic if \(\alpha=-1\). If \(\alpha=0\), the most singular term is instead the logarithmically divergent \( -\log{z} \); this term will also be included if \( \alpha\) is a negative integer, from integrating the \(1/u\) term.

\section{Conclusion}

We have arrived in quite a natural way at Wu's postulated form for the regulator mass \(M_{R}^{2}\), resolving the irreducible loop integrals into a simple closed-form expression involving the incomplete Gamma-function. The approach is reasonably general, and the dimension of the spacetime does not enter in a significant way. We hope that our closed-form expressions will both ease computation in the theory, and enable physicists to draw on a wider mathematical theory in using LORE.


\begin{thebibliography}{10}

\bibitem{ISI:000187232400003}
Y.~Wu, {\em {Int. J. Mod. Phys. A}} {\bf {18}}, 5363
  ({2003}).
  
\bibitem{Chapling:2016LORE}
R.~Chapling, {\em {Mod. Phys. Lett. A}}, {\bf {31}}, 1650030 (2016).

\bibitem{ISI:000224011700003}
Y.~Wu, {\em {Mod. Phys. Lett. A}} {\bf {19}}, 2191  ({2004}).



%
%
%
%
%
%

\bibitem{Ma:2006fk}
Y.-L. Ma and Y.-L. Wu, {\em Int. J. Mod. Phys. A} {\bf
  21}, 6383  (2006).

\bibitem{Dai:2005ve}
Y.-B. Dai and Y.-L. Wu, {\em Eur. Phys. J. C} {\bf 39}, 1  (2005).

\bibitem{ISI:000258778900002}
J.-W. Cui and Y.-L. Wu, {\em {Int. J. Mod. Phys. A}} {\bf
  {23}}, 2861  ({2008}).

\bibitem{ISI:000293182400002}
J.-W. Cui, Y.-L. Ma and Y.-L. Wu, {\em {Phys. Rev. D}} {\bf {84}}, 025020
  ({2011}).

\bibitem{Wu:2014uq}
Y.-L. Wu, {\em Int. J. Mod. Phys. A} {\bf 29},   1430007
  (2014).
  
\bibitem{GrimmettStirzaker:2001pr}
G.~R.~Grimmett and D.~R.~Stirzaker, {\em Probability and Random Processes}, (Oxford University Press, 2001)




%
%
%
%
%



\end{thebibliography}
\end{document}